\documentclass[aps,pre,10pt,floatfix,showpacs,superscriptaddress,longbibliograph,twocolumn]{revtex4-2}

\usepackage[T1]{fontenc}
\usepackage{amsmath}
\usepackage{amssymb}
\usepackage[english]{babel}
\usepackage{booktabs}
\usepackage{graphicx}
\usepackage[colorlinks=true,citecolor=blue,linkcolor=blue]{hyperref} 
\usepackage{algorithm}  
\usepackage{algpseudocode}
\usepackage{multirow}
\usepackage{newtxmath}
\usepackage{newtxtext}
\usepackage{subfigure}
\usepackage{lipsum}
\usepackage{physics}
\usepackage{siunitx}

\newcommand{\KLD}[2]{D_{\mathrm{KL}}\left(#1 \middle\| #2 \right)}
\newcommand{\s}{\mathbf{s}}
\newcommand{\dtheta}{\delta \theta}
\DeclareMathOperator*{\argmin}{arg\,min}

\begin{document}

\title{Efficient Optimization of Variational Autoregressive Networks with Natural Gradient}

\author{Jing Liu}
\email{jing.liu@itp.ac.cn}
\affiliation{CAS Key Laboratory for Theoretical Physics, Institute of Theoretical Physics, Chinese Academy of Sciences, Beijing 100190, China}

\author{Ying Tang}
\affiliation{Institute of Fundamental and Frontier Sciences, University of Electronic Science and Technology of China, Chengdu 611731, China}
\affiliation{Key Laboratory of Quantum Physics and Photonic Quantum Information, Ministry of Education, University of Electronic Science and Technology of China, Chengdu 611731, China}

\author{Pan Zhang}
\email{panzhang@itp.ac.cn}
\affiliation{CAS Key Laboratory for Theoretical Physics, Institute of Theoretical Physics, Chinese Academy of Sciences, Beijing 100190, China}
\affiliation{School of Fundamental Physics and Mathematical Sciences, Hangzhou Institute for Advanced Study, UCAS, Hangzhou 310024, China}
\date{\today}

\begin{abstract}

Estimating free energy is a fundamental problem in statistical mechanics.
Recently, machine-learning-based methods, particularly the variational autoregressive networks (VANs) have been proposed to minimize variational free energy and to approximate the Boltzmann distribution. 
VAN enjoys notable advantages, including the exact computation of the normalized joint distribution and fast sampling, which are critical features often missing in Markov chain Monte Carlo algorithms.
However, VAN also faces significant computational challenges.
These include difficulties in the optimization of variational free energy in a complicated parameter space and slow convergence of learning.
In this work, we introduce an optimization technique based on natural gradients to the VAN framework, namely \textit{ng-VAN}, to enhance the learning efficiency and accuracy of the conventional VAN.
The method has computational complexity cubic in the batch size rather than in the number of model parameters, hence it can be efficiently implemented for a large VAN model.
We carried out extensive numerical experiments on the Sherrington-Kirkpatrick model, spin glasses on random graphs, and the two-dimensional Ising model.
Our results indicate that compared with the conventional VAN, ng-VAN significantly improves the accuracy in estimating free energy and converges much faster with shorter learning time. 
This allows extending the VAN framework's applicability to challenging statistical mechanics problems that were previously not accessible.
\end{abstract}

\maketitle

\section{Introduction}
\label{section:intro}
In statistical mechanics, a fundamental problem is how to estimate the partition function, the statistical quantities, and obtain unbiased samples from the equilibrium Boltzmann distribution 
\begin{equation}
    p_{\mathrm{eq}}(\s) = \frac{1}{Z}e^{-\beta E(\s)}.
\end{equation}
Here $\s\in\{\pm 1\}^{N}$ denotes a configuration with $N$ spins, $\beta$ is the inverse temperature, $E(\s)$ is the energy for a configuration $\s$, and $Z=\sum_{\s}e^{-\beta E(\s)}$ denotes the partition function.

Recently, a prominent application of machine learning methods, \textit{variational autoregressive networks} (VANs)~\cite{wu2019solving,nicoli2020asymptotically,liu2021solving,pan2021solving}, have been introduced to solve statistical mechanics problems.
In VAN, an autoregressive neural network is employed as a variational ansatz $q_\theta(\s)$ to approximate the Boltzmann distribution ${p_{\mathrm{eq}}(\s)}$.
Autoregressive neural networks belong to the broader category of generative models in machine learning~\cite{tomczak2022deep} and have emerged as a powerful framework for modeling high-dimensional complex data distributions~\cite{lei2024generative}.
They utilize the chain rule of probabilities to parameterize the joint distribution of sequential data, such as text and images.
Notable examples include large language models trained on large corpora~\cite{vaswani2017attention,brown2020language} and image generation by PixelCNN~\cite{oord2016pixel,oord2016conditional}.
In contrast, the learning process of VAN is no longer based on maximum likelihood but is akin to reinforcement learning, with an objective of minimizing the Kullback-Leibler divergence $\KLD{q_\theta}{p_{\mathrm{eq}}}$, which necessitates sampling from the variational distribution itself.
The autoregressive characteristics of the model enable tractable likelihood and fast sampling~\cite{larochelle2011neural,germain2015made,oord2016pixel,oord2016conditional}, a significant advantage over traditional Markov chain Monte Carlo (MCMC) methods.
Moreover, VAN can provide an asymptotically unbiased estimator of physical observables~\cite{nicoli2020asymptotically}.
Likewise, autoregressive models have also been employed to represent variational wave functions in studies of variational Monte Carlo (VMC)~\cite{sharir2020deep,hibat-allah2020recurrent,melko2024language}.

Despite its advantages, VAN also faces considerable challenges, particularly regarding optimization difficulties associated with nonconvex parameter landscapes.
VAN typically relies on first-order gradient-based optimization algorithms.
Although state-of-the-art optimizers such as Adam~\cite{kingma2015adam} incorporate momentum information and show superior performances compared to traditional stochastic gradient descent~\cite{schmidt2021descending}, they may still encounter difficulties in high-dimensional parameter space and exhibit slow convergence speed.
This limitation necessitates the need for more advanced optimization methods.
One promising approach for solving this issue is the natural gradient descent~\cite{amari1998natural,pascanu2014revisiting,martens2020new}, which has already proven effective in reinforcement learning~\cite{kakade2001natural,peters2008natural} as well as in VMC scenarios~\cite{becca2017quantum,carleo2017solving}.
The fundamental concept behind the natural gradient is to account for the curvature of the loss function by employing a preconditioner to rescale the gradient updates.
Geometrically, the natural gradient represents the steepest descent direction in the model's distribution space rather than in the Euclidean parameter space.
However, this method requires the inversion of the Fisher information matrix, a computational complexity of $\mathcal{O}({N_p^3})$ with $N_p$ variational parameters.
In recent advancements in VMC~\cite{chen2024empowering,rende2024simple} and machine learning~\cite{chen2023efficient}, researchers have proposed efficient algorithms for implementing the natural gradient, especially in situations where the number of samples or batch size $N_b$ is significantly smaller than $N_p$.
Rather than inverting the original $N_p \times N_p$ matrix, it only requires the inversion of a significantly smaller $N_b \times N_b$ matrix which shares the same spectrum properties as the Fisher information matrix.
In this way, the natural gradient method can be applied to neural network models with up to $10^6$ parameters, which is far beyond traditional approaches.

In this work, we propose to integrate the natural gradient into the VAN framework, namely \textit{ng-VAN}, to enhance both learning efficiency and accuracy compared to the conventional VAN. 
In particular, we employ an efficient method for computing the inverse of the Fisher information matrix with computational complexity growing cubic in the batch size, rather than the number of parameters.
Our results demonstrate that the natural gradient method can significantly accelerate the convergence of VAN during training and achieve more accurate estimated variational free energy compared to first-order optimizers such as Adam. 
The rest of the paper is organized as follows: 
in Sec.~\ref{section:methods}, we first introduce VAN and its learning, followed by an introduction to the natural gradient method and how to implement it efficiently in the context of VAN learning.
In Sec.~\ref{section:results}, we present the experimental results, and finally, in Sec.~\ref{section:conclusion}, we summarize our conclusions and discuss the potential applications of our methods.

\section{Methods}
\label{section:methods}

\subsection{Variational autoregressive networks}

The VANs proposed in Ref.~\cite{wu2019solving} employ a variational ansatz for the target Boltzmann distribution, parameterized by a set of learnable parameters $\theta = \{\theta_i\}_{i=1}^{N_p}$, where $N_p$ denotes the total number of parameters.
This variational distribution, denoted as $q_\theta(\s)$, obeys the autoregressive property, i.e., can be decomposed into the product of conditional probabilities:
\begin{align}
    q_\theta(\s) = \prod_{i=1}^{N} f_\theta(s_i | s_1, \cdots, s_{i-1}),
\end{align}
where $N$ is the number of spins.
The primary objective of the training is to optimize these parameters in order to minimize the distance between the variational distribution $q_\theta(\s)$ and the true Boltzmann distribution $p_{\mathrm{eq}}(\s) = e^{-\beta E(\s)} / Z$, which is quantified by the Kullback-Leibler (KL) divergence metric $\KLD{q_\theta}{p_{\mathrm{eq}}}$.
Minimizing this KL divergence is equivalent to minimizing the variational free energy, defined as 
\begin{align}
    F_q(\theta) = \mathbb{E}_{\s \sim q_\theta} \left[ E(\s) + \frac{1}{\beta} \ln q_\theta(\s) \right].
\end{align}
Moreover, the variational free energy $F_q(\theta)$ serves as an upper bound for the exact free energy.
A crucial component required for gradient-based optimization is the gradient of the variational free energy with respect to $\theta$.
Denoting $R(\s) = E(\s) + (1/\beta) \ln q_\theta(\s)$, the gradient can be expressed as 
\begin{align}
    \eta = \nabla_\theta F_q(\theta) = \mathbb{E}_{\s \sim q_\theta} \left[ R(\s) \nabla_\theta \ln q_\theta(\s) \right].
    \label{eq:reinforce}
\end{align}
A baseline-corrected form can also be employed for variance reduction, given by $\eta = \mathbb{E}_{\s \sim q_\theta}[(R(\s) - b) \nabla_\theta \ln q_\theta(\s)]$, where $b=\mathbb{E}_{\s \sim q_\theta} [R(\s)]$ is a constant termed as baseline.
This gradient expression is known as the score function gradient estimator~\cite{mohamed2020monte}, also referred to as the REINFORCE algorithm~\cite{williams1992simple}, a specific instance of the policy gradient method from reinforcement learning.
Note that the gradient is $\nabla_\theta \ln q_\theta(\s)$ weighted by reward signal $R(\s)$, indicating that configurations with large rewards are penalized in the optimization process, hence reduce the variational free energy.

VANs have shown improved performance over traditional variational mean-field methods and have attracted significant research interest, including their integration with importance sampling~\cite{nicoli2020asymptotically}, MCMC methods~\cite{nicoli2020asymptotically,mcnaughton2020boosting,wu2021unbiased}, and their extension to larger sparse graphs~\cite{pan2021solving}.
However, as illustrated in Sec.~\ref{section:intro}, it also encounters several significant challenges.
In contrast to the case where a dataset is available for maximum likelihood estimation, in the training of VAN, gradient estimation in each iteration requires drawing samples from the current variational distribution.
Due to the sequential nature of autoregressive sampling, this process is highly time consuming.
Furthermore, gradients derived from samples based on a suboptimal variational distribution from the preceding step may not provide sufficient information for the subsequent optimization, leading to poor overall performance.

\subsection{Natural gradient descent}

Here we present a concise and self-contained derivation of the natural gradient within the framework of optimizing VAN.
The natural gradient can be derived by identifying the update $\dtheta$ that minimizes the variational free energy $F_q(\theta + \dtheta)$, subject to the constraint that the KL divergence between the current variational distribution $q_\theta$ and the updated distribution $q_{\theta + \dtheta}$ is equal to a specified value $\epsilon$.
Mathematically, this is formulated as
\begin{align}
    \dtheta^\ast = \argmin_{\KLD{q_\theta}{q_{\theta + \dtheta}} = \epsilon} F_q(\theta + \dtheta).
\end{align}
The Lagrange function for this optimization problem is given by
\begin{align}
    \mathcal{L}(\dtheta,\lambda) = F_q(\theta + \dtheta) + \lambda \left[ \KLD{q_\theta}{q_{\theta + \dtheta}} - \epsilon \right]. 
\end{align}
The KL divergence can be approximated by $ \KLD{q_\theta}{q_{\theta + \dtheta}} \approx (1/2) \dtheta^\top S \dtheta$ (see Appendix~\ref{appendix:proof} for details), where $S$ is the Fisher information matrix (FIM), defined as
\begin{align}
    S = \mathbb{E}_{\s \sim q_\theta} \left[\nabla_\theta \ln q_\theta(\s) \nabla_\theta \ln q_\theta(\s)^\top \right].
    \label{eq:fisher}
\end{align}
With $N_b$ samples drawn from $q_\theta$, the FIM is estimated as
\begin{align}
    S = \frac{1}{N_b} \sum_{i=1}^{N_b} \nabla_\theta \ln q_\theta(\s^{i}) \nabla_\theta \ln q_\theta(\s^{i})^\top,
\end{align}
where the superscript denotes the $i$-th sample.
By employing a first-order Taylor series expansion of the variational free energy, we obtain
\begin{align}
    \mathcal{L}(\dtheta,\lambda) = F_q(\theta) + \nabla_\theta F_q(\theta)^\top \dtheta + \frac{\lambda}{2} \dtheta^\top S \dtheta - \lambda \epsilon.
\end{align}
Setting the derivative of the Lagrangian with respect to $\dtheta$ to zero yields the natural gradient update formula:
\begin{align}
    \label{eq:natural-grad}
    \dtheta = - \frac{1}{\lambda} S^{-1} \nabla_\theta F_q(\theta),
\end{align}
where the factor $1 / \lambda$ can be integrated into the learning rate $\alpha$.
Furthermore, by substituting Eq.~(\ref{eq:natural-grad}) into the constraint $(1/2) \dtheta^\top S \dtheta = \epsilon$, the adaptive learning rate $\alpha$ can be determined as
\begin{align}
    \label{eq:adaptive-lr}
    \alpha = \frac{1}{\lambda} = \sqrt{\frac{2\epsilon}{\nabla_\theta F_q(\theta)^\top S^{-1} \nabla_\theta F_q(\theta)}}.
\end{align}
The resulting adaptive learning rate ensures that the parameter updates are appropriately scaled.

Suppose that we choose the Euclidean distance in the parameter space as the constraint: $\norm{\dtheta}_2^2 = \epsilon$, the same derivation leads to the conventional gradient descent algorithm, i.e., the FIM is equivalent to a trivial identity matrix (see Appendix~\ref{appendix:proof} for derivations).
Intuitively, the steepest descent in the distribution space is more appropriate, as a minor change in the parameter space may result in significant changes in the probability distributions before and after the update.
By employing the inverse of the Fisher information matrix, $S^{-1}$, as a gradient preconditioner,  natural gradient updates can effectively account for the local curvature of the manifold, leading to more efficient and stable optimization.
However, the issue associated with the natural gradient method lies in the necessity of inverting the FIM, a procedure with a computational complexity of $\mathcal{O}(N_p^3)$.
This cubic complexity can become prohibitively expensive as the model size increases, particularly for deep neural network architectures.
Next, we will introduce an efficient algorithm for resolving this issue.

\subsection{Efficient algorithm for implementing natural gradient descent}

In a recent work, Chen and Heyl introduced a novel method called minSR~\cite{chen2024empowering}, which enables the training of neural-network quantum states~\cite{carleo2017solving} with up to $10^6$ parameters using stochastic reconfiguration, a quantum generalization of natural gradient method.
Then Rende \textit{et al.} reformulated this approach using a linear algebra identity~\cite{rende2024simple}. 
Specifically, in the context of VAN, Eq.~(\ref{eq:reinforce}) can be estimated using $N_b$ samples drawn from the variational distribution:
\begin{align}
    \eta &= \mathbb{E}_{\s \sim q_\theta} \left[ R(\s) \nabla_\theta \ln q_\theta(\s) \right] \nonumber\\
    &= \frac{1}{N_b} \sum_{i=1}^{N_b} R(\s^i) \nabla_\theta \ln q_\theta(\s^i),~~\mathrm{where}~\s^i \sim q_\theta(\s)
    \label{eq:reinforce_samples}
\end{align}
By defining $O_{i k} = (1/ \sqrt{N_b})\partial \ln q_\theta(\s^i) / \partial \theta_k$ and $R_i = (1 / \sqrt{N_b}) R(\s^i)$, where $O$ is a $N_b \times N_p$ matrix and $R$ is a $N_b$-dimensional vector, Eq.~(\ref{eq:reinforce_samples}) can be rewritten in matrix notation as
\begin{align}
    \eta = O^\top R.
\end{align}
We consider the condition of over-parameterization, i.e., $N_p \gg N_b$, which is typical in the implementation of VAN.
Using the above notation, the empirical FIM from Eq.~(\ref{eq:fisher}) can be written as 
\begin{align}
    S = O^\top O,
\end{align}
allowing Eq.~(\ref{eq:natural-grad}) to be represented as
\begin{align}
    \dtheta = -\alpha (O^\top O)^{-1} O^{\top} R. 
    \label{eq:SR}
\end{align}
By leveraging a linear algebra identity $(O^\top O)^{-1} O^{\top} = O^\top (O O^\top)^{-1} $, we have
\begin{align}
    \dtheta = -\alpha O^\top (O O^\top)^{-1} R.
    \label{eq:minsr_solve}
\end{align}
If we use a fixed learning rate $\alpha$, the computational complexity of Eq.~(\ref{eq:minsr_solve}) is $\mathcal{O}(N_b^3 + N_p N_b^2)$ determined by the matrix inversion of $O O^\top$ and the matrix multiplication of $O^\top (O O^\top)^{-1} R$.
In contrast to directly inverting the FIM ($\mathcal{O}(N_p^3)$), our algorithm is faster especially when $N_p \gg N_b$.
Besides, the computational complexity for Eq.~(\ref{eq:adaptive-lr}) is $\mathcal{O}(N_p N_b)$ instead of $\mathcal{O}(N_p^2)$, determined by the dot product between $\dtheta$ and $O^\top R$.

It is noteworthy that Eq.~(\ref{eq:minsr_solve}) is also known as natural gradient with the Moore-Penrose pseudo-inverse and has been extensively studied in the machine learning literature~\cite{thomas2014genga, bernacchia2018exact, zhang2019fast, cai2019gramgaussnewton, karakida2020understanding}.
In the implementation, $O O^\top$ can be ill-conditioned and it requires regularization scheme.
Usually a non-negative damping factor $\xi$ is introduced $(O O^\top + \xi I)^{-1}$ for numerical stability.
See Ref.~\cite{schmitt2022jvmc} for more details.
Furthermore, in addition to the linear algebra trick utilized in Ref.~\cite{rende2024simple}, there are alternative techniques for efficiently solving the inverse of the FIM, such as the algorithm based on Cholesky decomposition proposed in Ref.~\cite{chen2023efficient}.

\section{Results}
\label{section:results}

\begin{figure*}[!htb]
    \centering
    \includegraphics[width=\linewidth]{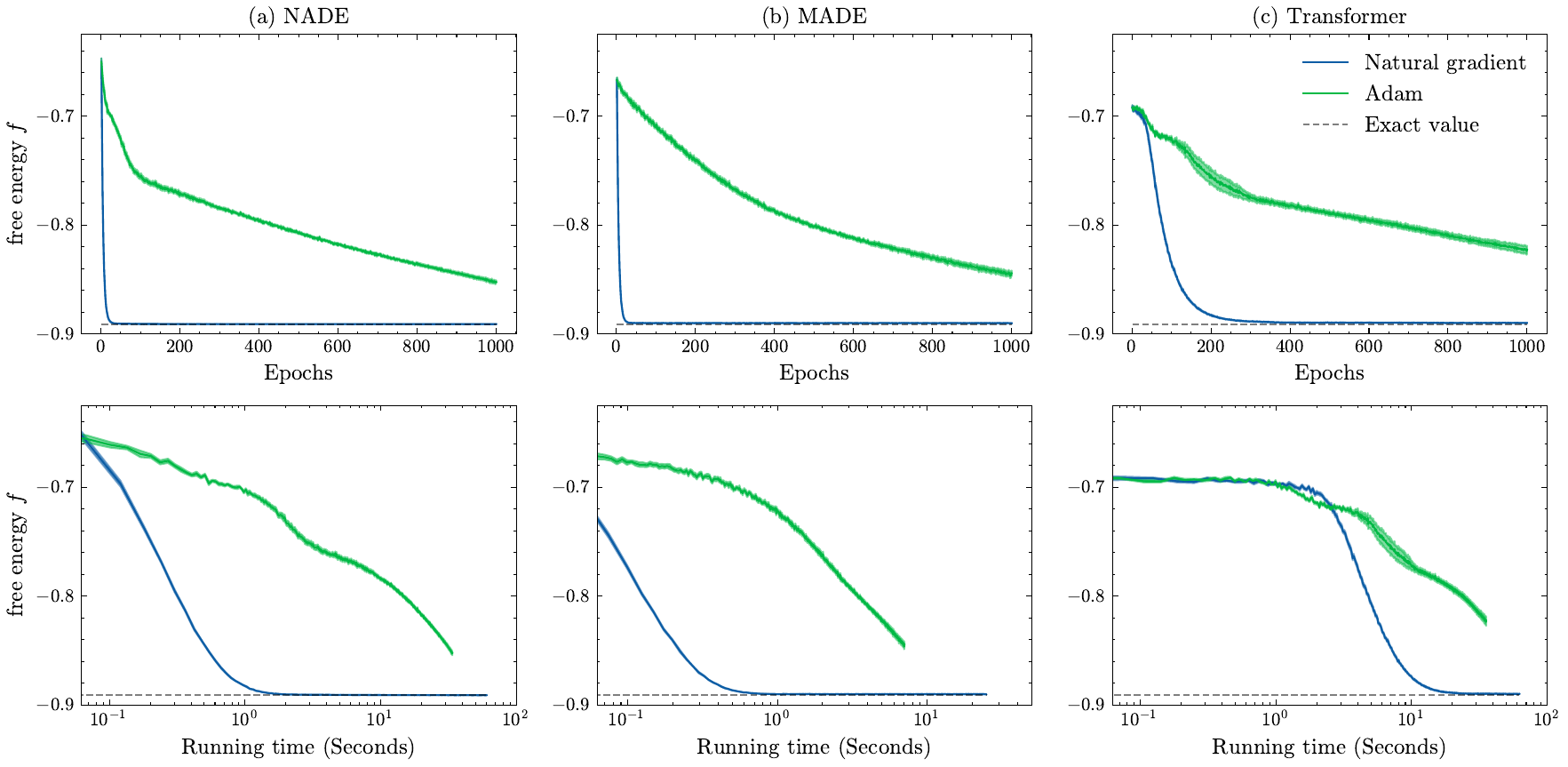}
    \caption{For the Sherrington-Kirkpatrick (SK) model with $N=30$ spins at the spin-glass phase transition point $\beta=1.0$, we compare the evolution of the variational free energy as training epochs and training time obtained by Adam optimizer (conventional VAN) and natural gradient (ng-VAN), using NADE, MADE, and the transformer model as autoregressive models. 
    The results are averaged over 10 independent parameter initialization on the same SK instance.
    }
    \label{fig:converge}
\end{figure*}

\begin{figure*}[!htb]
    \centering
    \includegraphics[width=\linewidth]{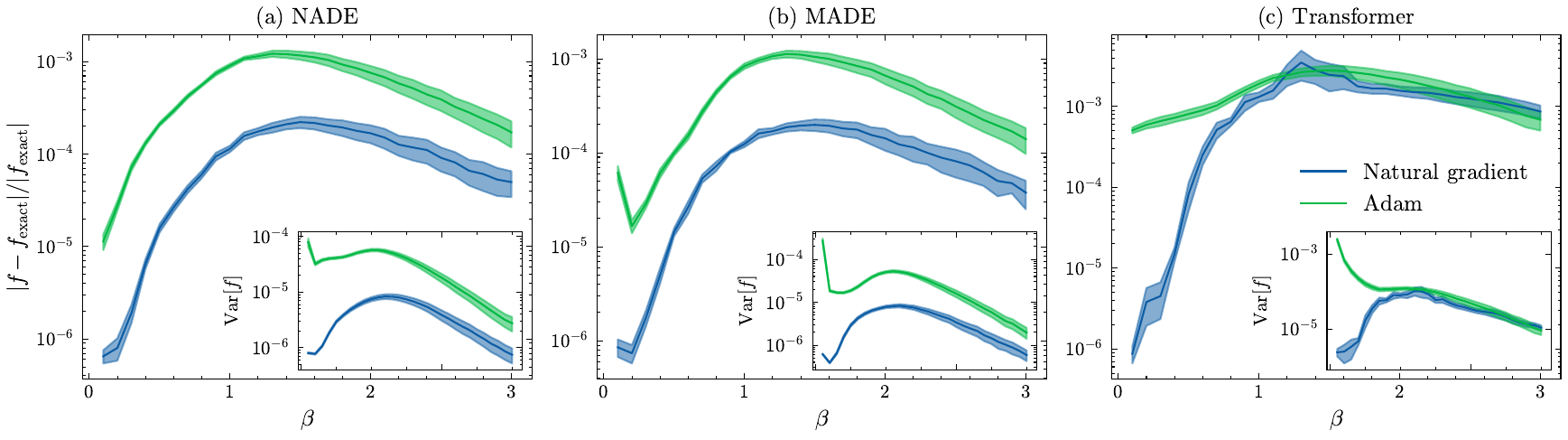}
    \caption{For the SK model with $N=30$ spins, the relative error between the exact and variational results as a function of inverse temperature $\beta$ is presented for Adam optimizer (conventional VAN) and natural gradient (ng-VAN), using NADE, MADE, and the transformer.
    The inset plots show the variance of the variational free energy, which also measures the training quality.
    The results are averaged over 10 independent SK instances with different couplings.}
    \label{fig:sk}
\end{figure*}

In our experiments, we utilized the Neural Autoregressive Distribution Estimator (NADE)~\cite{larochelle2011neural}, the Masked Autoencoder for Distribution Estimation (MADE)~\cite{germain2015made}, a decoder-only transformer~\cite{vaswani2017attention} and PixelCNN~\cite{oord2016pixel,oord2016conditional} as our variational neural network ans\"atze. 
A concise overview of these models is provided in Appendix~\ref{appendix:nn}.
We implemented the natural gradient method using PyTorch~\cite{paszke2019pytorch} and tested it on an NVIDIA GPU.
The hyperparameters of our experiments are summarized in Appendix~\ref{appendix:hyperparams}.
We performed a series of experiments on the well-established Sherrington-Kirkpatrick (SK) model~\cite{sherrington1975solvable}, spin glasses defined on the random regular graph (RRG) as well as the two-dimensional Ising model.
In the SK model, the couplings $J_{ij}=J_{ji}$ are drawn from a Gaussian distribution with a mean of zero and a variance of $1 / \sqrt{N}$.
For the RRG model, we set the degree to be $d=3$, and the coupling $J_{ij}$ are randomly sampled from $\{-1, +1\}$ with equal probability.

\subsection{Performance on variational free energy}

In the first study, in order to investigate the convergence characteristics of the natural gradient method, we performed an experiment on an SK instance with $N=30$ spins, where we can still calculate the exact values of free energy through brute-force enumeration. 
After initializing the parameters, we employed both the natural gradient descent and Adam optimizer to minimize the variational free energy directly at $\beta=1.0$, the spin glass phase transition point.
From the first row of Fig.~\ref{fig:converge}, it is evident that when employing the Adam optimizer, neither NADE, MADE, nor the transformer achieved convergence of the variational free energy, even after 1000 epochs.
In contrast, when we switched to the natural gradient method, all three models successfully converged to the exact free energy value, particularly for NADE and MADE, which showed convergence in fewer than 100 epochs.
In the second row of Fig.~\ref{fig:converge}, we also present the evolution of the variational free energy over running time. 
Although the training of the natural gradient method takes longer than Adam under the same epochs, the natural gradient can converge in a shorter overall duration, and provides an exponential speed up in computational time in this case.
This experiment illustrates that the natural gradient method has a substantial advantage in accelerating the convergence rate during training as compared to the Adam optimizer.

Then we followed the training procedure described in Ref.~\cite{wu2019solving}, implementing a temperature annealing process that starts at a high temperature.
This approach facilitates the training of neural networks while effectively mitigating the potential occurrence of mode collapse phenomena.
Figure~\ref{fig:sk} shows the relative error of the variational free energy for NADE, MADE and transformer, compared to the exact values.
Notably, the results obtained by the natural gradient method demonstrate a superiority over those achieved by the Adam optimizer.
We also observed that the performance of the transformer was inferior to that of NADE and MADE, regardless of whether the Adam optimizer or the natural gradient method was employed. 
The spectra of the FIM, as discussed in Ref.~\cite{donatella2023dynamics}, shows that the FIM trace for the transformer model increases more rapidly during training compared to that of MADE and NADE.
This phenomenon is associated with steeper curvature region of the loss function, suggesting  that training the transformer in the VAN framework is more challenging and require more careful tuning of hyperparameters.
Additionally, we provide inset plots showing the variance of the variational free energy, which equals zero when $q_\theta = p_{\mathrm{eq}}$ ($p_{\mathrm{eq}}=e^{-\beta E(\s)}/Z$, assuming that mode collapse does not occur, otherwise, $Z$ may not present the original partition function of the Boltzmann distribution), referred to as the zero variance condition.
In this way, the variance serves as an indicator of the proximity between the variational distribution and the true Boltzmann distribution, providing a measure of the training quality~\cite{nicoli2021estimation}.
From Fig.~\ref{fig:sk}, it is evident that the variance associated with the natural gradient method is consistently lower than that of the Adam optimizer.

\begin{figure}[!htb]
    \centering
    \includegraphics[width=\linewidth]{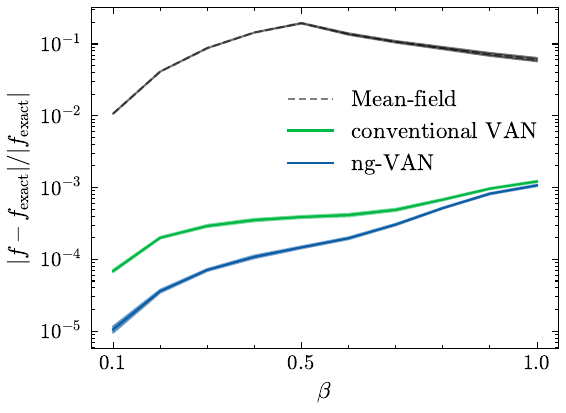}
    \caption{For the random regular graph (RRG) with $N=256$ spins and degree $d=3$, we compare the relative error of the variational free energy between MADE using Adam optimizer (conventional VAN) and MADE using natural gradient (ng-VAN). The results of the mean-field ansatz is also presented (dashed line).
    The results are averaged over 10 independent RRG instances.}
    \label{fig:rrg}
\end{figure}

\begin{figure*}[!htb]
    \centering
    \includegraphics[width=\linewidth]{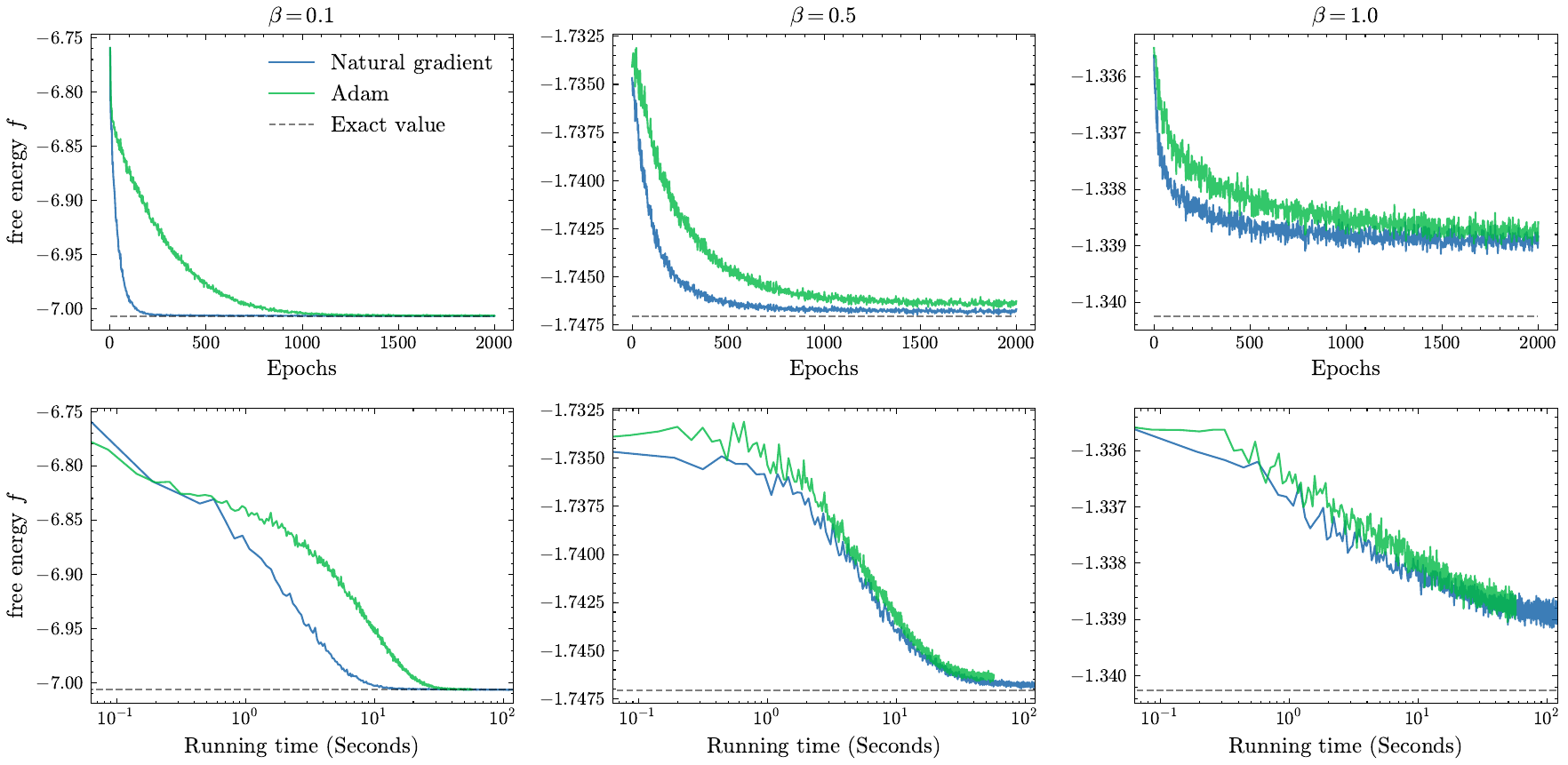}
    \caption{Evolution of the variational free energy and the variance obtained by MADE, for the random regular graph with $N=256$ spins and $d=3$.}
    \label{fig:rrg-loss-evolution}
\end{figure*}

\begin{figure}[!htb]
    \centering
    \includegraphics[width=\linewidth]{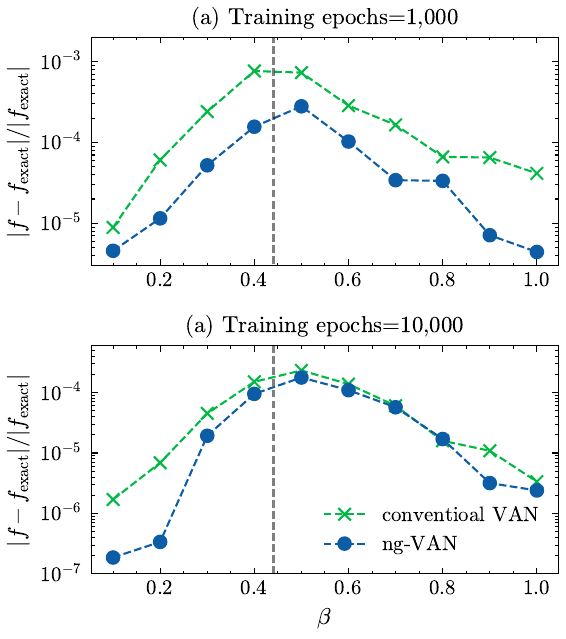}
    \caption{The relative error of the variational free energy to the exact solution of ferromagnetic Ising model on $16 \times 16$ lattice with open boundary condition obtained by natural gradient (ng-VAN) and Adam optimizer (conventional VAN). The dashed line represents the critical temperature $\beta_c=\ln(1+\sqrt{2})/2$.}
    \label{fig:ising2d}
\end{figure}

\begin{figure*}[!htb]
    \centering
    \includegraphics[width=\linewidth]{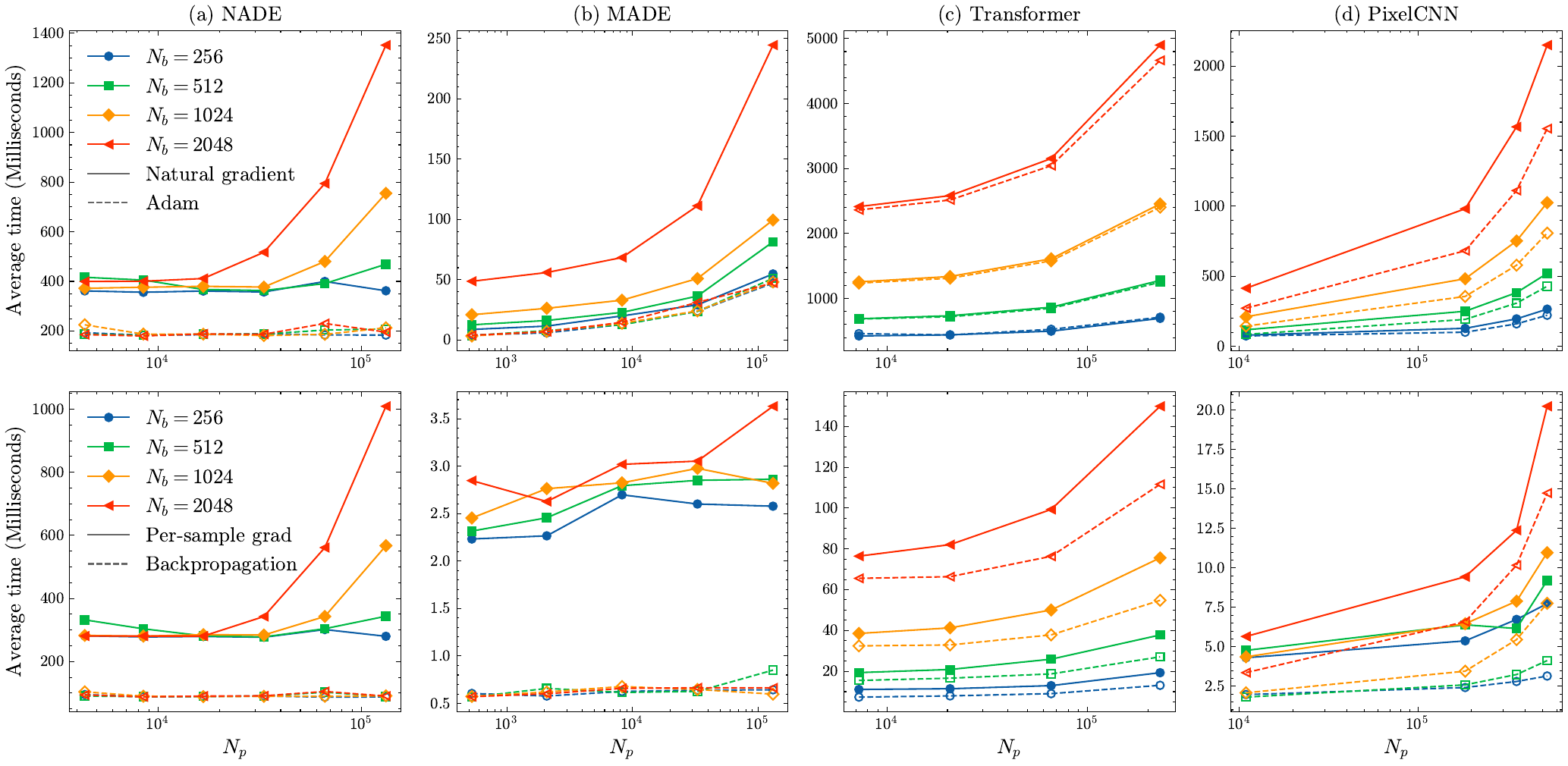}
    \caption{Top row: the average computational cost per epoch under different number of parameters ($N_p$) and batch size ($N_b$) for NADE, MADE, transformer, and PixelCNN. Here the differences in running time are attributed to the fact that natural gradient methods require the computation of per-sample gradients rather than backpropagation, along with the necessity of solving Eq.~(\ref{eq:minsr_solve}).
    Bottom row: the average computational cost of backpropagation and the computation of per-sample gradients.}
    \label{fig:time}
\end{figure*}

\begin{figure}[!htb]
    \centering
    \includegraphics[width=\linewidth]{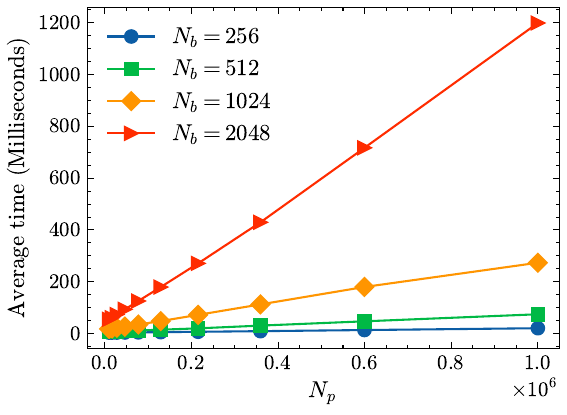}
    \caption{The average computational cost under a different number of parameters ($N_p$) and batch size ($N_b$) to solve the inverse of the Fisher information matrix.}
    \label{fig:time-benchmark}
\end{figure}

We further carried out experiments on large sparse networks.
For RRG with $N=256$, the free energy can still be obtained through exact tensor network contraction~\cite{gray2021hyperoptimized,liu2021tropical,pan2022simulation,pan2022solving}.
Figure~\ref{fig:rrg} presents the relative error of the variational free energy of MADE.
In Fig.~\ref{fig:rrg-loss-evolution}, we also plot the evolution of the loss for one instance.
The results clearly indicate that the natural gradient method exhibits a faster convergence speed compared to the Adam optimizer, while also yielding lower variational free energies across different temperature regions.
It is important to note that, at first glance, the advantage of the natural gradient appear to diminish as the temperature decreases.
This is primarily attributable to the annealing schedule employed during training, which helps to mitigate mode collapse.
The effective performance of Adam at low temperatures depends on this annealing strategy.
In contrast, if training is initiated directly at low temperatures, Adam is susceptible to being trapped in local optima and may experience mode collapse, while the natural gradient method exhibits faster convergence and remains more effective compared to Adam, see Appendix~\ref{appendix:hyperparams} for more results.
Consequently, for ground state and combinatorial optimization problems with Ising formulations~\cite{lucas2014ising}, the natural gradient method allows for initiation from lower temperatures and requires fewer annealing steps.

Finally, we consider the two-dimensional ferromagnetic Ising model.
We employ open boundary condition without external fields, allowing for the exact solution of the free energy to be derived using the Kac-Ward formula~\cite{kac1952combinatorial}.
We utilized the PixelCNN architecture~\cite{oord2016pixel,oord2016conditional} and also follow the experimental setting outlined in Ref.~\cite{wu2019solving}.
In Fig.~\ref{fig:ising2d}, we present the relative error of the variational free energy across different temperatures for system of $16 \times 16$.
When the training epochs are set to \num{1000}, the natural gradient exhibits significant advantages over the Adam optimizer. 
Furthermore, even with an increased training epochs of \num{10000}, the natural gradient continues to outperform Adam.

\subsection{Computational overhead of natural gradient}

Here, we investigate the computational overhead associated with the natural gradient method. 
Each epoch in the training of VAN comprises three primary steps: (1) sampling from $q_\theta(\s)$, (2) executing the forward pass [calculating the variational free energy], and (3) performing backpropagation for evaluating the gradient with respect to the model parameters.
In the case of the natural gradient method, the backpropagation is substituted with the computation of per-sample gradients, i.e., the matrix $O$ in Eq.~(\ref{eq:SR}).
This is a nontrivial task since deep learning frameworks usually accumulate the gradients on the entire batch of samples directly.
Besides, solving Eq.~(\ref{eq:minsr_solve}) requires additional computational costs in double-precision floating-point format.

In Fig.~\ref{fig:time}, we present the average computational cost per epoch for both the natural gradient method and the Adam optimizer under different number of parameters ($N_p$) and batch size ($N_b$).
We also provide the running time for the computation of per-sample gradients and executing backpropagation.
We observe that, for NADE, increases in $N_p$ and $N_b$ do not significantly affect the computational time for Adam optimizer. 
Furthermore, its calculation of per-sample gradients is notably more time consuming compared to other models.
This is due to the fact that NADE requires $N$ forward passes to evaluate $q_\theta(\s)$, similar to recurrent neural networks, while MADE, transformer, and PixelCNN necessitate only a single forward pass.
Consequently, the computational inefficiency of NADE leads to lower GPU utilization as well as a more time-intensive process for computing per-sample gradients.
For MADE, transformer, and PixelCNN, the time difference between backpropagation and computation of per-sample gradients is marginal, so the extra time required by the natural gradient method is primarily attributable to the matrix inversion process.
Figure~\ref{fig:time-benchmark} depicts the average computational cost associated with solving the inverse of the FIM, which scales linearly with $N_p$ as expected.

\section{Conclusions and outlooks}
\label{section:conclusion}

In this work, we have successfully integrated an efficient computational approach for the natural gradient method into the VAN framework~\cite{wu2019solving}.
Our approach, referred to as ng-VAN, drawing on recent developments in machine learning and VMC~\cite{chen2024empowering,rende2024simple}, utilizes a linear algebra identity to solve the inverse of the Fisher information matrix with computational cost cubic in the batch size.
This enables us to train neural networks with up to $10^6$ parameters, which is beyond the capabilities of traditional approaches.

Our experiments on various spin glasses benchmarks suggest that compared to conventional VAN, the ng-VAN significantly improves convergence speed. 
Furthermore, ng-VAN also enhances the accuracy of estimating variational free energy and other relevant physical quantities.
While the natural gradient method offers these advantages, it requires additional computational costs compared to first-order gradient-based algorithms such as Adam~\cite{kingma2015adam}.
This is primarily due to the necessity of calculating per-sample gradients as well as the inverting of the Fisher information matrix.
Consequently, this leads to a trade-off between accelerating convergence speed and the increased computational time associated with each iteration.
We also acknowledge the existence of other approximation methods for the Fisher information matrix, such as the Kronecker-Factored Approximate Curvature~\cite{martens2015optimizing}, which have been extensively investigated in VMC studies~\cite{drissi2024secondordera}.

We highlight that our approach is applicable to the optimization of tensor network generative models~\cite{han2018unsupervised,cheng2019tree,liu2023tensor}, variational neural annealing~\cite{hibat-allah2021variational}, neural-enhanced Monte Carlo sampling techniques~\cite{nicoli2020asymptotically,mcnaughton2020boosting,ciarella2023machinelearningassisted}, and the recently proposed TwoBo architecture~\cite{biazzo2023autoregressive,biazzo2024sparse}.
Besides, ng-VAN can also be employed to investigate out-of-equilibrium systems, including chemical reaction networks~\cite{tang2023neuralnetwork,liu2024distilling} and nonequilibrium lattice models~\cite{tang2024learning}.

Our PyTorch implementation is available in Ref.~\cite{src}.

\begin{acknowledgments}
We thank C. Ye for helpful discussions and F. Pan for kindly providing the exact free energy values of RRG.
J.L. acknowledges support from Project No. 12405047 of the National Natural Science Foundation of China. 
Y.T. acknowledges support from Projects No. 12322501 and No. 12105014 of the National Natural Science Foundation of China. 
P.Z. acknowledges support from Projects No. 12325501, No. 12047503, and No. 12247104 of the National Natural Science Foundation of China, and Project No. ZDRW-XX-2022-3-02 of Chinese Academy of Sciences. 
Part of the computation is carried out at the HPC Cluster of ITP-CAS.
\end{acknowledgments}

\appendix
\section{Proofs and derivations in the main text}
\label{appendix:proof}

First, we give the derivation of the KL divergence approximation formula.
Using a second-order Taylor expansion, the KL divergence can be expressed as follows, with $\theta' = \theta + \dtheta$:
\begin{align}
    &\KLD{q_\theta}{q_{\theta'}} \nonumber \\
    \approx& \KLD{q_\theta}{q_\theta} + \dtheta^\top \eval{\nabla_{\theta'} \KLD{q_\theta}{q_{\theta'}}}_{\theta'=\theta} 
    \nonumber \\
    +& \frac{1}{2} \dtheta^\top \eval{\nabla_{\theta'}^2 \KLD{q_\theta}{q_{\theta'}}}_{\theta'=\theta} \dtheta \nonumber \\
    =& - \dtheta^\top \mathbb{E}_{q_\theta} \left[ \nabla_\theta \ln q_\theta \right] - \frac{1}{2} \dtheta^\top \mathbb{E}_{q_\theta} \left[\nabla_\theta^2 \ln q_\theta \right] \dtheta.
\end{align}
The first term is equal to zero since
\begin{align}
    &\mathbb{E}_{q_\theta} \left[ \nabla_\theta \ln q_\theta \right] \nonumber \\=& \sum_{\s} q_\theta(\s) \frac{1}{q_\theta(\s)} \nabla_\theta q_\theta(\s) = \nabla_\theta \sum_{\s}q_\theta(\s) = 0,
\end{align}
and the Fisher information matrix $S$ is the negative expected Hessian matrix of the log-likelihood:
\begin{align}
    &-\mathbb{E}_{q_\theta} \left[\nabla_\theta^2 \ln q_\theta \right]_{ij} \nonumber \\ =& -\mathbb{E}_{q_\theta}\left[ \pdv{\ln q_\theta}{\theta_i}{\theta_j}\right] \nonumber \\
    =& -\mathbb{E}_{q_\theta} \left[ \pdv{\theta_j}(\frac{1}{q_\theta}\pdv{q_\theta}{\theta_i})\right] \nonumber \\
    =& -\mathbb{E}_{q_\theta} \left[ \frac{1}{q_\theta}\pdv{q_\theta}{\theta_i}{\theta_j} \right] + \mathbb{E}_{q_\theta} \left[\pdv{\ln q_\theta}{\theta_i} \pdv{\ln q_\theta}{\theta_j} \right] \nonumber \\
    =& S_{ij}.
\end{align}
Thus, the KL divergence can be approximated as:
\begin{align}
    \KLD{q_\theta}{q_{\theta + \dtheta}} \approx \frac{1}{2}\dtheta^\top S \dtheta.
\end{align}

If we choose the Euclidean distance in the parameter space as the constraint: $\norm{\dtheta}^2_2=\epsilon$, it leads to conventional gradient descent method.
To illustrate this, we can reformulate the problem as
\begin{align}
    \dtheta^\ast = \argmin_{\norm{\dtheta}^2_2=\epsilon} F_q(\theta + \dtheta).
\end{align}
Note that the squared norm $\norm{\dtheta}^2_2$ can be expressed as $\dtheta^\top \dtheta$. 
The corresponding Lagrangian for this optimization problem is
\begin{align}
    \mathcal{L}(\dtheta,\lambda) &= F_q(\theta + \dtheta) + \lambda \left(\norm{\dtheta}^2_2 - \epsilon \right) \nonumber\\
    &= F_q(\theta) + \nabla_\theta F_q(\theta)^\top \dtheta + \lambda \dtheta^\top \dtheta - \lambda \epsilon.
\end{align}
By setting the derivative of the Lagrangian with respect to $\dtheta$ to zero, we obtain
\begin{align}
    \dtheta &= -\frac{1}{\lambda} \nabla_\theta F_q(\theta),
\end{align}
which corresponds to the conventional gradient descent method.
In this case, there is no preconditioner to rescale the gradients except for the learning rate factor $1/\lambda$, in other words, the FIM is an identity matrix.

In the main text, we utilize the linear algebra identity: $(O^\top O)^{-1} O^{\top} = O^\top (O O^\top)^{-1}$.
To establish the aforementioned identity, we begin with the fundamental matrix identity $I = (O O^\top) (O O^\top)^{-1}$.
By multiplying both sides by $O^\top$, we obtain
\begin{align}
    O^\top = O^\top (O O^\top) (O O^\top)^{-1}=(O^\top O) O^\top (O O^\top)^{-1}.
\end{align}
Subsequently, multiplying both sides by $(O^\top O)^{-1}$ yields
\begin{align}
    (O^\top O)^{-1} O^{\top} = O^\top (O O^\top)^{-1}.
\end{align}
Similarly, we can also prove a similar result in the presence of a damping term, specifically: 
\begin{align}
    (O^\top O + \xi I_{N_p \times N_p})^{-1} O^{\top} = O^\top (O O^\top + \xi I_{N_b \times N_b})^{-1}.
\end{align}

\section{Variational neural-network ans\"atze}
\label{appendix:nn}

Here, we give a brief overview of NADE~\cite{larochelle2011neural}, MADE~\cite{germain2015made}, decoder-only transformer~\cite{vaswani2017attention}, and PixelCNN~\cite{oord2016pixel,oord2016conditional}.
The NADE parameterizes the conditional probability as follows:
\begin{align}
    h_i &= \sigma\left(W_{:, :i} \mathbf{x}_{:i} + c\right), \\
    p(x_i=1|x_1,\cdots,x_{i-1}) &=\sigma\left(V_{i, :} h_i + b_i \right).
\end{align}
Here we use Python's slicing and indexing syntax.
To obtain the spin configuration, we apply the transformation $\s = (1 + \mathbf{x}) / 2$ to convert the binary variable $\mathbf{x}$ into spin configuration.
The trainable parameters of NADE include $\{W \in \mathbb{R}^{H \times N}, c \in \mathbb{R}^H, V \in \mathbb{R}^{N \times H}, b \in \mathbb{R}^{N}\}$, with $N$ and $H$ denoting the system size and hidden dimension respectively.

For MADE, in the simplest form, it reduces to the fully-visible sigmoid belief network:
\begin{align}
    p(x_i|x_1,\cdots,x_{i-1}) = \sigma \left(\sum_{j<i} W_{i,j} \mathbf{x}_j + b_i\right),
\end{align}
where the trainable parameters are $\{W \in \mathbb{R}^{N \times N}, b \in \mathbb{R}^{N}\}$.
For a general MADE, it has $\mathcal{O}(N^2)$ parameters compared to $\mathcal{O}(NH)$ of NADE.

In the decoder-only transformer, the input configurations and their positional information are first mapped through a linear transformation into an embedding matrix $Q \in \mathbb{R}^{N \times d_k}$, where $d_k$ is the embedding dimension. 
The causal self-attention mechanism is then defined as follows:
\begin{align}
    \mathrm{CausalSelfAttention}(Q) = \mathrm{Softmax}(\frac{QQ^{\top}}{\sqrt{d_k}} + M)Q,
\end{align}
where $M$ is a masking matrix that assigns negative infinity for positions corresponding to future spins, thereby enforcing the autoregressive property.
Following the attention mechanism, the output is processed through a feed-forward network (FFN), which consists of two linear transformations separated by a nonlinear activation function, typically ReLU:
\begin{align}
    \mathrm{FFN}(x) = \mathrm{ReLU}(x W_1 + b_1) W_2 + b_2.
\end{align}
The causal self-attention and FNN architecture can be stacked into a transformer block, supplemented by residual connections and layer normalization.
Typically, a transformer model contains several such blocks improve its representational capacity. 
Finally, the output from the FNN is fed into a softmax layer, converting the model's outputs into conditional probabilities.
Additionally, we can transform the spin configurations into a sequence of fixed-size patches, thereby reducing the dimensionality of the input~\cite{dosovitskiy2021image}.

PixelCNN employs a convolutional neural network architecture to model the joint distribution of pixel intensities in an image, generating new images sequentially, one pixel at a time~\cite{oord2016pixel,oord2016conditional}. 
By leveraging masked convolutions, PixelCNN ensures that the prediction for each pixel is influenced only by previously generated pixels, maintaining the autoregressive property. 
PixelCNN has been shown to outperform VAN that do not utilize convolutional networks on the two-dimensional lattice systems~\cite{wu2019solving}.

\section{Hyperparameters and implementation details}
\label{appendix:hyperparams}

In our experiments, we used the default setting of Adam optimizer~\cite{kingma2015adam} with a learning rate of $10^{-3}$.
For the natural gradient method, we found that selecting a fixed learning rate $\alpha_{\mathrm{fixed}}$ or setting $\epsilon$ in Eq.~(\ref{eq:adaptive-lr}) within the range of $[0.01, 1]$ generally yields good performance; however, the fixed learning rate tends to produce better results.
If the learning rate for natural gradient optimization is excessively small, convergence can be significantly slower.
Additionally, a learning rate schedulers can also be employed, such as decaying $\alpha_{\mathrm{fixed}}$ or $\epsilon$ over several epochs, to further enhance performance.
Regarding the damping factor $\xi$ for regularization, we observed that setting it within a reasonable range, such as $[10^{-4},10^{-2}]$, has minimal impact on overall performance.
Therefore, in our experiments, we consistently set $\xi$ to $10^{-3}$.

In the experiment on the SK model with $N=30$, we set the hidden dimension $H=64$ for NADE.
For MADE, we employ a two-layer architecture with a hidden dimension of 150.
The transformer model only consists of a single transformer block, with embedding dimension $d_k=32$, FNN dimension $d_{ff}=128$ and number of heads $n_{\mathrm{head}}=4$.
In this way, these three models exhibit a comparable number of parameters.
During the annealing procedure, the training starts at $\beta=0.1$ and then subsequently decreased in increments of $\delta \beta=0.1$ until reaching $\beta=3.0$.
In other words, model parameters at a given $\beta$ are initialized using a well-trained model from the previous step at $\beta - \delta \beta$.
The batch size is 1024 and the training epochs at each temperature are set to 1000.
In the experiment on the RRG with $N=256$, we utilize a single-layer MADE.
Compared to the experiments on the SK model, the training epochs are increased to 2000, and the learning rate is set to $0.05$ for the natural gradient.
As for the two-dimensional Ising model, we initiate training from $\beta=0$ and gradually increase $\beta$ over $N_{\mathrm{annealing}}$ steps (training epochs) until the target temperature is achieved. 
We also adopted the PixelCNN architecture used in Ref.~\cite{wu2019solving}, which comprises \num{714113} learnable parameters.
We set $N_{\mathrm{annealing}}$ to $\num{1000}$ and $\num{10000}$, with a learning rate of $0.1$ for the natural gradient.

\begin{figure*}[!htb]
    \centering
    \includegraphics[width=\linewidth]{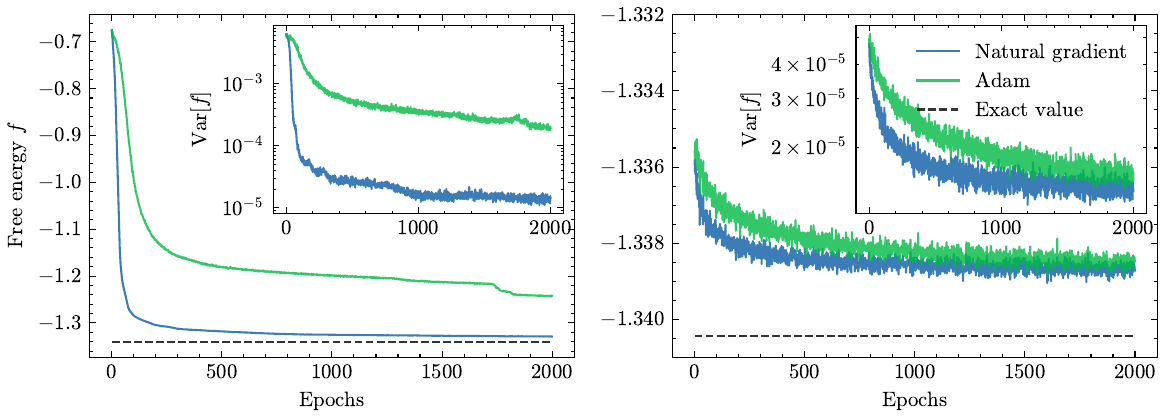}
    \caption{Evolution of the variational free energy and the variance obtained by MADE, for the random regular graph with $N=256$ spins and $d=3$. Left: training the model directly at $\beta = 1.0$. Right: training the model from $\beta=0.1$ and gradually increase the inverse temperature in increments of $\delta \beta=0.1$ to $\beta=1.0$.}
    \label{fig:rrg_appendix}
\end{figure*}

It is important to emphasize that the annealing schedule described above plays a critical role in the training process, particularly for the Adam optimizer.
At each inverse temperature $\beta$, the model parameters are initialized using a well-trained model from the preceding step at $\beta - \delta \beta$.
This annealing schedule ensures that the Adam optimizer remains effective even at higher values of $\beta$.
As illustrated in Fig.~\ref{fig:rrg_appendix}, for an RRG with $N=256$, training directly at $\beta=1.0$ (as opposed to starting at $\beta=0.1$ and then gradually increase the inverse temperature) demonstrates that natural gradient optimization maintains robust performance.
In contrast, the Adam optimizer exhibits significantly slower convergence, further highlighting the superior efficiency of natural gradient methods in regimes of low temperature.

\bibliography{refs.bib}

\end{document}